\documentclass[pra,twocolumn,superscriptaddress,showpacs,amsmath,amssymb]{revtex4-1}

\usepackage{graphicx}
\usepackage{latexsym}
\usepackage{amsmath}
\usepackage{amssymb}
\usepackage{amsfonts}
\usepackage{color}

\usepackage{amssymb}
\usepackage{bm}
\usepackage{verbatim}

\usepackage[colorlinks,bookmarks=false,citecolor=blue,linkcolor=red,urlcolor=blue]{hyperref}
\usepackage{color}
\usepackage[all]{hypcap} 

\newcommand{\rev}[1]{{\color{red}#1}}

\begin{document}

\title{Electronic structure of mesoscopic superconducting disk: \\
Quasiparticle tunneling between the giant vortex core and disk edge}

\author{A.~V.~Samokhvalov}
\affiliation{Institute for Physics of Microstructures, Russian Academy of Sciences, Nizhny Novgorod, Russia}
\affiliation{Lobachevsky State University of Nizhni Novgorod, 23 Prospekt Gagarina, 603950 Nizhni Novgorod, Russia}
\author{I.~A.~Shereshevskii}
\affiliation{Institute for Physics of Microstructures, Russian Academy of Sciences, Nizhny Novgorod, Russia}
\affiliation{Lobachevsky State University of Nizhni Novgorod, 23 Prospekt Gagarina, 603950 Nizhni Novgorod, Russia}
\author{N.~K.~Vdovicheva}
\affiliation{Institute for Physics of Microstructures, Russian Academy of Sciences, Nizhny Novgorod, Russia}
\author{M.~Taupin}
\affiliation{Institute of Solid State Physics, Vienna University of Technology, Wiedner Hauptstrasse 8-10, 1040 Vienna, Austria}
\author{I.~M.~Khaymovich}
\affiliation{Max Planck Institute for the Physics of Complex Systems, N\"{o}thnitzer Str. 38, 01187 Dresden}
\affiliation{Institute for Physics of Microstructures, Russian Academy of Sciences, Nizhny Novgorod, Russia}
\author{J.~P.~Pekola}
\affiliation{QTF Centre of Excellence, Department of Applied Physics, Aalto~University~School of Science, P.O. Box 13500, 00076 Aalto, Finland} \author{A.~S.~Mel'nikov}
\affiliation{Institute for Physics of Microstructures, Russian Academy of Sciences, Nizhny Novgorod, Russia}
\affiliation{Lobachevsky State University of Nizhni Novgorod, 23 Prospekt Gagarina, 603950 Nizhni Novgorod, Russia}

\begin{abstract}
The electronic structure of the giant vortex states in a
mesoscopic superconducting disk is studied in a dirty limit using
the Usadel approach. The local density of states profiles are shown to be strongly affected by the effect of quasiparticle (QP) tunneling between
the states localized in the vortex core and the ones bound to the sample edge. Decreasing temperature leads to a crossover
between the edge-dominated and core-dominated regimes in the magnetic field dependence of the tunneling conductance. This crossover is discussed
in the context of the efficiency of quasiparticle cooling by the magnetic field induced QP traps in various mesoscopic superconducting devices.
\end{abstract}
%

\maketitle

\section{Introduction}\label{IntroSection}
Vortex states in mesoscopic superconducting (SC) systems of the size
comparable to the superconducting coherence length,
have been well studied over the past few decades, mainly with the emphasis on
the dependence of the vortex configuration on the size and geometry of the sample.
In such nanoscale samples theory predicts that only a few vortices can be placed, and confinement
effects result in different exotic vortex configurations
unlike the triangular Abrikosov lattice
\cite{fink,Moshchalkov-Nature95,Peeters-Nature97,geim,Buisson,Peeters-PRL98,palacios,bruy,jadallah,Moshchalkov-Nature00,Melnikov-Nature02,Melnikov-PRB02,Peeters-PRL06}.
These exotic
configurations are formed by the interplay between
imposed boundary conditions and the repulsive
interactions between vortices.

The most remarkable consequence of this interplay
is the formation of the so-called
giant vortex state or multiquantum vortex when all the vortices merge in the disk center
 predicted mostly within the Ginzburg-Landau formalism provided the disk size is of order of the coherence length.
A variety of experimental methods have been used to verify these theoretical predictions:
(i)~Hall probe microscopy
\cite{Peeters-Nature97,geim,Moshchalkov-PRB04,Moshchalkov-PRB08},
(ii)~Bitter decoration~\cite{Grigorieva},
(iii)~scanning SQUID microscopy
\cite{Kokubo-PRB10},
(iv)~different tunneling experiments including scanning tunneling microscopy/spectroscopy studies
\cite{Peeters-PRL04,Roditchev-PRL11,Roditchev-PRL09,Nishio-PRL08,Moshchalkov-PRB16}.

The latter experimental approach is known to be sensitive to the electronic structure of the sample, namely,
to the local density of states of quasiparticle excitations and, thus, the
phenomenological Ginzburg~--~Landau theory often appears to be insufficient for the interpretation
of the experimental data. This
clear demand to the microscopic theory has stimulated theoretical activity in the field
concentrated mainly on the calculations based on the Bogolubov-de-Gennes theory
\cite{Melnikov-Nature02,Melnikov-PRB02,tanaka1,tanaka2,rainer,virt,esch,tanaka3,Melnikov-PRL,mrs1,mrs2,Peeters-PRL12},
i.e., on the clean limit corresponding to the
very large mean free path $\ell$ well exceeding both the coherence length $\xi_0$ and
 the sample size.
Certainly, the predictions made within such approach may be difficult to use for most of the experimentally available
samples for which the dirty limit conditions ($\ell\ll\xi_0$) are much more appropriate.
In particular, it is natural to expect that all the density of states features associated, e.g., with the
different anomalous spectral branches \cite{volovik} in the giant vortex or with the mesoscopic oscillations of the
Caroli-de Gennes-Matricon energy levels ~\cite{Melnikov-PRL} due to the finite sample size
should be smeared by disorder.
An adequate theoretical description of the sample electronic structure in this diffusive regime should be, of course,
based on the Usadel-type theory. And indeed such calculations are known to provide
an excellent tool for the analysis of the Abrikosov vortex lattices in unrestricted geometries
(see, e.g., \cite{Golubov-PRL94}). For multiquantum giant vortices these results have been generalized in
Ref.~\cite{Silaev-JPCM13} without accounting the effect of the sample boundary.

It is important to note that the demand in the theoretical explanation of the available data of scanning tunneling microscopy and spectroscopy (STM/STS) on the exotic vortex structures in mesoscopic samples~(see, e.g.,~\cite{NS_vort_PRL,NS_vort_JETP}) is
not the only motivation for the continuing research work in the field.
Nowadays, superconducting nanostructures have become an important element in
designing devices for
rapidly expanding fields of quantum computing, quantum memory, superconducting logic,
and metrology and they are obviously the main building blocks for the superconducting electronics.
However, superconductors are known to be easily poisoned by non-equilibrium quasiparticles and these extra excitations drastically affect the performance of above-mentioned quantum devices, e.g. via overheating or unwanted population in general.
To suppress overheating in a superconductor different types of
quasiparticle traps are used (see, e.g.,
Refs.~\cite{Nguyen-NJP13,Ullom-APL98,Khaymovich-NatCom16} and
references therein). One of the possible types of quasiparticle traps can be formed by
regions with the reduced superconducting gap which appear in the Meissner and
vortex states and can be successfully controlled by
the external magnetic field (see~\cite{Peltonen2011,Nsanzineza2014,Wang2014,Vool2014,Woerkom2015,Khaymovich-NatCom16,Nakamura17}).
Further progress in the field requires a quantitative theoretical description of both types of quasiparticle traps based on the vortex penetration
as well as on Meissner currents flowing mostly at the sample edge.
Thus, the main goal of our work is
to analyze the behavior of the local density of states in giant vortices penetrating to a circular superconducting sample of order of the coherence length, with proper accounting of the sample edge effects.
This analysis, to our mind, should provide an important step on the route to rather general model of quasiparticle traps in mesoscopic samples.

\begin{figure}[t!]
\includegraphics[width=0.4\textwidth]{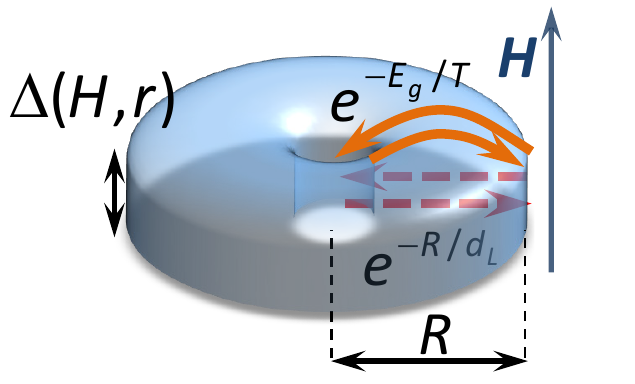}
\caption{(Color online) Schematic picture of the spatial order parameter distribution (shown by semi-transparent blue color)
in the superconducting disk of the radius $R$ with the giant $L$-fold vortex in the applied perpendicular magnetic field~$H$.
The exponential factor $e^{-R/d_L}$ ($e^{-E_{\rm g}/T}$) close to the red dashed (orange solid) lines corresponds to
the amplitude of the quantum tunneling (thermally activated) process.
} \label{Fig0}
\end{figure}

To elucidate the key results of our study it is useful to note that both the giant vortex cores and the sample edge with the flowing Meissner
screening currents can be clearly viewed as Andreev potential wells for quasiparticles in the clean limit~\cite{Melnikov-Nature02}.
On the other hand, the impurity scattering in the dirty limit surely modifies some spectral characteristics of these wells compared to the clean regime:
(i)~scattering broadens the
discrete levels of the Caroli-de Gennes-Matricon energy branch~\cite{CdGM}, which crosses the Fermi level,
suppressing the minigap in the spectrum \cite{Golubov-PRL94},
(ii)~scattering can also result in the
increase of minigap in the quasiparticle spectrum $E_{\rm g}$ at the sample edge because the changes in the quasiparticle momentum directions partially suppress the
effect of the Doppler shift of the quasiparticle energy in the presence of the surface currents \cite{eg1,eg2,eg3,eg4}.
The overall spectral characteristics and local density of states of the mesoscopic sample can be considered as an interplay
of the subgap states,
located in the vortices and in the regions with the reduced spectral gap $E_{\rm g}$ by Meissner currents, especially at the sample edge.
To illustrate this interplay we consider for instance different contributions to the zero bias conductance (ZBC) at the sample edge (see Fig.~\ref{Fig0}), which can be experimentally accessed in tunneling transport measurements.
The contribution of the giant vortex core states to this quantity can be estimated as follows:
$\sim \exp (-R/d_L)$, where $R$ is the distance from the vortex center to the boundary and $d_L$ is the effective decay length dependent on the vorticity $L$. For $L=1$ the latter length $d_1\simeq \xi_0$ is of order of the coherence length $\xi_0$.
The contribution of the edge states should include the temperature activation exponent $\sim \exp (-E_{\rm g}/T)$ due to the finite spectral minigap $E_{\rm g}$. These two terms are comparable for
a characteristic temperature $T^*(R) \simeq E_{\rm g} d_L/R$.
Thus, we conclude that in a sample of certain size $R$ for the temperatures larger than $T^*(R)$ the core contribution
is negligible at the sample edge and, consequently, the finite temperature masks the coupling of the Andreev wells in the vortex core and at the edge.
In the opposite limit of small temperatures $T<T^*(R)$, quantum-mechanical tunneling of the subgap quasiparticles between the vortex and the edge traps becomes observable in the experimentally measurable quantities and dominating over thermally-activated processes.
Here and further the Boltzmann's constant is set to unity, $k_{\rm B} = 1$.

The above estimates give us a simple criterion of the interplay of the core and edge state contributions which will be quantitatively confirmed by further calculations of the local density of states (LDOS)
in a diffusive mesoscopic SC disk in a wide interval of magnetic fields,
applied perpendicular to the sample plane. Note that these estimates can be of course applied not only for a vortex in a finite size sample
but also for any experimental geometry with vortices positioned close to the superconductor edge (see, e.g., STM images in Ref.~\cite{guill}).

 The paper is organized as follows. In
Sec.~\ref{ModelSection} we briefly discuss the basic equations. In
Sec.~\ref{CritTemp} we calculate the superconducting critical
temperature $T_{\rm c}$ and study the switching between the states with
different vorticity $L$ while sweeping the magnetic field. In
Sec.~\ref{LDOS} we find both analytically and numerically the
spatially resolved LDOS and study the behavior of the jumps in ZBC which are attributed to the entrance of a vortex into the
disk. We summarize our results in Sec.~\ref{SumUp}.

\section{Model and basic equations}\label{ModelSection}
Hereafter we consider a thin superconducting disk of a finite
radius $R$ of order of the coherence length at the
temperature $T$, placed in external magnetic field $\mathbf{H} = H
\mathbf{z}_0$ oriented perpendicular to the plane of the disk (Fig.~\ref{Fig0}). The
disk thickness is assumed to be small compared to the London penetration depth,
thus, the effective magnetic field penetration depth is
large. This allows us to neglect the contributions to the magnetic
field from supercurrents and, thus, $\mathrm{rot}\mathbf{A} =
\mathbf{B} \equiv \mathbf{H}$.
Using the notations $\tau^{-1}$ for the electron elastic scattering
rate and $T_{\rm cs}$ for the bare superconductor transition temperature
the dirty limit conditions can be written as $T_{\rm cs} \tau \ll 1$.
In this regime the normal ($\mathcal{G}$) and anomalous ($\mathcal{F}$)
quasiclassical Green's functions are described by the Usadel
equations \cite{Usadel-PRL70}, which are valid for the whole
temperature and magnetic field range.
Focusing on the axisymmetric multiquantum vortex states with
the vortex core positioned in the center of the disk $r=0$
\begin{equation}
 \Delta(\mathbf{r}) = \Delta_L(r)\, \mathrm{e}^{i L \varphi}\,,
\end{equation}
we consider solutions, homogeneous along the $z$-axis and characterized by a
certain angular momentum $L$, referred further as vorticity
\begin{equation}\label{eq:3}
 \mathcal{F}(\mathbf{r},\omega_n) = \mathcal{F}_L(r, \omega_n)\, \mathrm{e}^{i L \varphi }\,.
\end{equation}
Here we choose the cylindrical coordinate system $(r, \varphi, z)$
and the gauge $\mathbf{A}=(0,\, A_\varphi,\, 0)$, $A_\varphi =
r H / 2$.
Due to the symmetry of Usadel equations $\mathcal{F}$ is an
even function of $\omega_n$, $\mathcal{F}(r,-\omega_n) = \mathcal{F}(r,\omega_n)$,
so that it is enough to treat only positive $\omega_n$ values. In the
standard trigonometrical parametrization $\mathcal{G} =\cos\theta_L$, $\mathcal{F} =
\sin\theta_L\, \mathrm{e}^{i L \varphi}$, $\mathcal{F}^\dag =
\sin\theta_L\, \mathrm{e}^{-i L \varphi}$ the Usadel equations take the form
\begin{multline}\label{eq:1a}
 -\frac{\hbar D}{2} \left[\ \frac{1}{r}\,\frac{d}{d r} \left( r\,\frac{d \theta_L}{d r} \right)
 - \left( \frac{L - \phi_r }{r} \right)^2 \sin\theta_L\,\cos\theta_L\, \right] \\
 +\, \omega_n\, \sin\theta_L = \Delta_L(r)\, \cos\theta_L \ .
\end{multline}
The self-consistency equation for the singlet superconducting order parameter function reads
\begin{equation}\label{eq:2a}
 \frac{\Delta_L(r)}{g}
 - 2\pi T \sum_{n\geq0}\sin\theta_L\, = 0\ .
\end{equation}
Here $D =v_{\rm F} l / 3$ is the diffusion coefficient, $\Phi_0 = \pi \hbar c/e$ is the
flux quantum, $\omega_n = \pi T ( 2 n + 1 )$ is the Matsubara
frequency at the temperature $T$, $\phi_r = \pi r^2 H / \Phi_0$ is a dimensionless flux of the external
magnetic field $\mathbf{H}$ threading the circle of certain radius $r$, and the pairing parameter $g$ determines the bare critical temperature $T_{\rm cs}$ as
\begin{equation}\label{eq:interaction_param}
 \frac{1}{g} =
 \sum_{n=0}^{\Omega_{\rm D}/(2\pi T_{\rm cs})} \frac{1}{n + 1/2}\simeq\ln[\Omega_{\rm D}/2\pi T_{\rm cs}]+2\ln2 + \gamma \ ,
\end{equation}
with the Debye frequency $\Omega_{\rm D}$ and the Euler~--~Mascheroni constant $\gamma \simeq 0.5772$.
The coherence length $\xi_0 = \sqrt{\hbar D / 2 \Delta_0}$
plays the role of a typical lengthscale in
the Usadel equations.

The equations
(\ref{eq:1a}, \ref{eq:2a}) should be supplemented with the
boundary conditions at the disk edge $r = R$:
\begin{equation}\label{eq:4}
 \left.\frac{d \Delta_L} {d r}\right|_{R} = \left.\frac{d \theta_L} {d r}\right|_{R} = 0\,.
\end{equation}

\section{Critical temperature of superconducting transitions with different vorticities}\label{CritTemp}
For the temperatures close to
the critical temperature of the superconducting transition $T\lesssim T_{\rm c}(H)$,
we can restrict ourselves by the solution of the Usadel equations Eqs.~(\ref{eq:1a}, \ref{eq:2a}) linearized
in the anomalous Green function ($\sin\theta_L \simeq \theta_L$):
\begin{multline}
 -\frac{\hbar D}{2} \left[\ \frac{1}{r}\,\frac{d}{d r} \left( r\,\frac{d \theta_L}{d r} \right)
 - \left( \frac{L - \phi_r }{r} \right)^2 \theta_L\, \right] \label{eq:1b} \\
 +\, \omega_n\, \theta_L = \Delta_L(r)\,,
\end{multline}
\begin{equation}\label{eq:2b}
 \frac{\Delta_L(r)}{g}
 - 2\pi T \sum_{n\geq0}\theta_L\, = 0\ .
\end{equation}
In these linearized equations the relation between the anomalous Green
function $\theta_L(r)$ and the order parameter $\Delta_L(r)$ can be
written in the standard form
\begin{equation} \label{eq:6}
 \theta_L(r,\omega_n) = \frac{\Delta_L(r)}{\omega_n + \Omega_L}\,,
\end{equation}
where $\Omega_L$ is the depairing parameter depending on the disk
radius $R$ and the external magnetic field ${\bf H}$. Thus, the
solution of Eq.~(\ref{eq:1b}) in the region $r \le R$ can
be expressed via the confluent hypergeometric function of the
first kind (Kummer's function $K(a,b,z)$
\cite{Abramowitz-Handbook})
\begin{subequations}\label{eq:4ab}
\begin{align}
\label{eq:4a}
\Delta_L(r) &= \theta_L(r)(\omega_n+\Omega_L) = C_L\, f_L(\phi_r)\,, \\
 f_L(\phi_r) &= \mathrm{e}^{-\phi_r/2}
 \phi_r^{\vert\,L\,\vert/2}\,K\left(\,a_L,\,b_L,\,\phi_r\,\right)\,.\label{eq:4b}
\end{align}
\end{subequations}
Here $C_L$ is a constant,
and the parameters $a_L$ and $b_L$
depend on the vorticity $L$ as follows (see Appendix~\ref{App1}
for details):
$$
 a_L = \frac{1}{2}\, \left( | L | - L + 1
 - \frac{\Phi_0 \Omega_L}{\pi \hbar D H} \right)\,,
 \; b_L = | L | + 1 \,.
$$
The boundary condition (\ref{eq:4}) for the orbital mode $L$ written in the
form (\ref{eq:4ab}) results in the following algebraic equation
\begin{multline} \label{eq:3bc}
 \Gamma_L(a_L,\,\phi) = b_L\,\left(\, | L | - \phi\, \right)\, K(a_L,\,b_L,\,\phi) \\
 +\, 2 \phi\, a_L\,K(a_L+1,\,b_L+1,\,\phi) = 0\,.
\end{multline}
The equation (\ref{eq:3bc}) determines the implicit dependence of
the parameter $a_L$ on the flux $\phi(H) = \pi R^2 H / \Phi_0\equiv H/H_0 $ through the disk for a fixed value of vorticity
$L$ normalized to the flux quantum. Here we introduced a characteristic field $H_0 = \Phi_0 / \pi R^2$.

The solutions $a_L^{(n)}$ of the equation
(\ref{eq:3bc}) give a set of values $\Omega_L^{(n)}$ which depend
on the normalized flux threading the whole disk $\phi$ and the
disk radius $R$: $\Omega_L=\Omega_L(\phi,\,R)$. Finally,
substituting the expression (\ref{eq:6}) into the
self-consistency condition (\ref{eq:2b}) one obtains
the following 
equation for the critical temperature
$T_L$ of the state with the vorticity $L$:
\begin{equation}\label{eq:2c}
 \ln \frac{T_L}{T_{\rm cs}} = \Psi\left( \frac{1}{2} \right) - \Psi\left( \frac{1}{2} +
 \frac{\Omega_L}{2 \pi T_L} \right)\,,
\end{equation}
where $\Psi$ is the digamma function.
%
\begin{figure}[t!]
\includegraphics[width=0.45\textwidth]{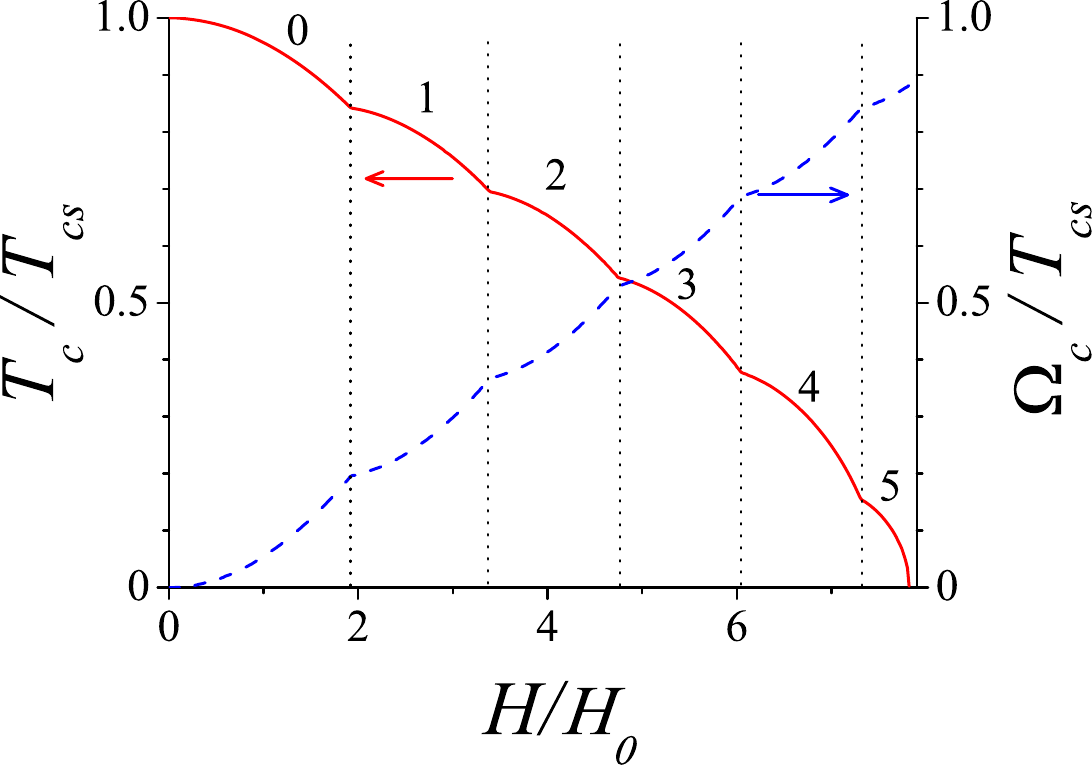}
\caption{(Color online) The dependence of the critical temperature
$T_{\rm c}$ (solid red line) and the depairing parameter $\Omega_{\rm c}$ (dashed blue
line) on the external magnetic field.
Here we choose $R = 4 \xi_0$ .
The numbers near the curves denote the corresponding
values of the vorticity $L_{\rm c}$. The dotted vertical
lines correspond to the fluxes $\phi = \phi_L$, where the switching of
the orbital modes $L \rightleftarrows L+1$ takes place. } \label{Fig1}
\end{figure}
%
In accordance with the self-consistency equation (\ref{eq:2c}),
the minimal value of the depairing parameter
\begin{equation}\label{eq:2d}
 \Omega_{\rm c} = \underset{L,\,n}{\rm min}
 \left\{\, \Omega_L^{(n)}(\phi,\,R)\, \right\}
\end{equation}
determines the vorticity $L_{\rm c}$ and the critical temperature $T_{\rm c} =
T_{L_{\rm c}}$ of the orbital mode, which nucleates in the disk of the radius
$R$ placed in the external magnetic $\mathbf{H}$.

Figure~\ref{Fig1} shows typical dependencies of the critical
temperature $T_{\rm c}$ and the depairing parameter $\Omega_{\rm c}$ on the
external magnetic flux $\phi$ across the disk for a fixed value of
the disk radius $R$. The phase boundary $T_{\rm c}(\phi)$ exhibits an oscillatory behavior similar to the
well-known
Little-Parks oscillations \cite{Little-PRL62,Parks-PRA64}, caused
by the transitions between the states with different angular momenta
$L$. The values of the normalized
flux through the disk $\phi_L$, where the switching of the orbital
modes $L \rightleftarrows L+1$ takes place, obey the equations
\begin{equation} \label{eq:3d}
 \Gamma_L(a_L,\,\phi_L) = 0,\, \quad \Gamma_{L+1}(a_{L+1} ,\,\phi_L) = 0\,,
\end{equation}
and do not depend on the disk radius $R$: $\phi_L \simeq
1.92;\, 3.40;\, 4.74;\, 6.04;\, 7.30; \ldots $ for $L = 0 \div 5
\ldots$. The magnetic field of the switching between modes $L$ and $L+1$
is determined by the expression $H_{\rm s} = H_0 \phi_L$.
The values of the dimensionless fluxes corresponding to the vorticity switching coincide with the ones found
in Ref.~\cite{jadallah} for a superconducting disk within the Ginzburg--Landau theory.
This coincidence comes from the obvious fact that the linearized Usadel equation
(\ref{eq:1b}) after the substitution of the expression (\ref{eq:6}) becomes similar to the linearized Ginzburg--Landau
equation. Surely, this similarity does not extend to the full behavior of the $T_L(H)$ curve
determined by Eq.~(\ref{eq:2c}). Note also that both the depairing factor and, thus, the critical temperature depend strongly
on the disk radius $R$: $\Omega_L \sim R^{-2} \tilde\Omega_L (\phi)$, where $\tilde\Omega_L(\phi)$ is a certain function of the dimensionless flux $\phi$ only.
One can see that the decrease in the $R$ value results in the decrease in the number of observable different vortex
states.

\section{Density of States}\label{LDOS}
\begin{figure}
\includegraphics[width=0.45\textwidth]{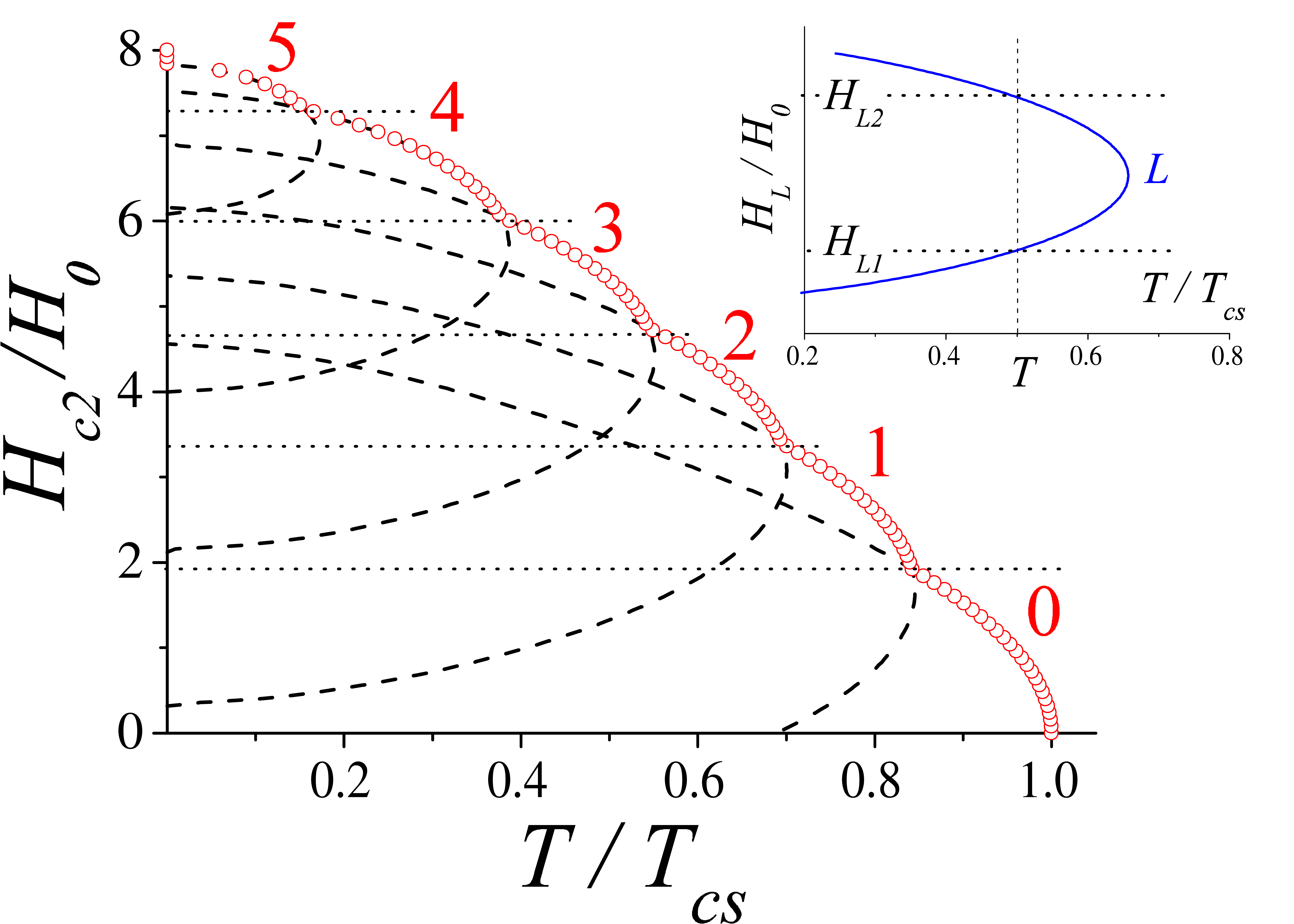}
\caption{(Color online) Schematic temperature dependence of the
upper critical field $H_{\rm c2}$ of superconducting phase transition
of $2D$ disks. Dashed lines show the dependence of the critical
magnetic fields $H_{L}(T)$ for the orbital mode with the vorticity
$L$ explicitly written in the plot. The dotted horizontal lines
are given by the relation $\phi(H) = \phi_L$ corresponding to the switching between the
orbital modes. The inset shows a typical dependence of
the critical magnetic fields $H_{L}(T)$ for the orbital mode
$L$.}\label{Fig2-Phic}
\end{figure}

For a fixed temperature $T$ the orbital mode $L$ exists in the
interval of the magnetic field values $0 \le H_{L1} \le H \le
H_{L2}$ which satisfy the condition $T_L(\phi(H)) \ge T$ (see
the inset in Fig.~\ref{Fig2-Phic}). Figure~\ref{Fig2-Phic} shows a
typical temperature dependence of the upper critical field for the
disk,
$$
 H_{\rm c2} = \underset{L}{\rm max} \left\{\,
 H_{L}(T)\,\right\}\,,
$$
affected by the transitions between different orbital states.
%
In order to analyze the characteristics of the sample far from the phase transition line
we return back to the nonlinear Usadel theory and consider full free energy functional:
\begin{multline}\label{usadelfull}
 F_L = 2 \pi T \sum\limits_{\omega_n < \Omega_{\rm D}} \int\limits_{0}^{R}
 r\,d r \bigg\{ \left(\frac{\partial\theta_L}{\partial r}\right)^2 + \left(\frac{L-\phi_r}{r}\right)^2\, \sin^2\theta_L \bigg. \\
 \bigg. - \frac{4}{\hbar D}\left(\omega_n \cos\theta_L + \Delta_L \sin\theta_L\right) \bigg\}
 + \frac{2}{\hbar D}\int\limits_{0}^{R} r\,d r\, \frac{\Delta_L^2}{g} \,.
\end{multline}
Focusing now on the effect of the switching between different vortex states on the density of states we should note that
experimentally this quantity can be most directly probed by the measurements of the local differential conductance:
\begin{eqnarray}
 G_L(\varepsilon\,,r\,,\phi) &=& \frac{d I/d V}{\left(d I/d V\right)_N} \label{eq:16} \\
 &=& \int\limits_{-\infty}^{\infty} d\varepsilon\, %
 \frac{N_L(\varepsilon\,,r\,,\phi)}{N_0}
 \frac{\partial f(\varepsilon-eV)}{\partial V}\,, \nonumber
\end{eqnarray}
where $V$ is the applied bias voltage, $(dI/dV)_N$ is a conductance of
the normal metal junction, and $f(\varepsilon) = 1/\left(1 +
\exp(\varepsilon/T )\right)$ is the Fermi function.

\subsection{High magnetic field: $H \lesssim H_{\rm c2}$}\label{LDOS-Tc}
We start our analysis from the limit of high magnetic fields close to the phase transition line $H_{\rm c2}(T)$ shown in
Fig.~\ref{Fig2-Phic} when
the solution of the Usadel equations can be significantly simplified due to the smallness of
the functions $\Delta_L$ and $\theta_L$.
In this case one can use the solution of the linearized theory (\ref{eq:6}, \ref{eq:4ab}). 
The constant $C_L$ in Eq.~(\ref{eq:4a})
should be found from the nonlinear Usadel theory (\ref{eq:1a}).
For this purpose we write the corresponding free energy up to the fourth power of $\Delta_L$ and $\theta_L$:
\begin{multline}\label{eq:14a}
 \frac{\hbar D}{2}\left(F_L-F_{\rm N}\right) = \int\limits_{0}^{R} r\,d r\, \bigg\{\frac{\Delta_L^2}{g}
 -2 \pi T \sum\limits_{\omega_n<\Omega_{\rm D}}\Delta_L \theta_L \\
 -2 \pi T \sum\limits_{\omega_n<\Omega_{\rm D}}\bigg[\frac{\hbar D}{2}\left(\frac{L-\phi_r}{r}\right)^2\, \frac{\theta_L^4}{3} + \omega_n \frac{\theta_L^4}{12} - \frac{\Delta_L\theta_L^3}{3} \bigg]\bigg\}
 \,,
\end{multline}
where $F_{\rm N}$ is the free energy of the normal state.
Using the above self-consistency equation for $T_{\rm c}(H)$
\begin{equation}
 \frac{1}{g} = \sum_{\omega_{n{\rm c}}<\Omega_{\rm D}}\frac{2\pi T_{\rm c}(H)}{\omega_{n{\rm c}}+\Omega_L}\,
\end{equation}
with $\omega_{n{\rm c}} = \pi T_{\rm c}(H) ( 2 n + 1 )$, and the
relation (\ref{eq:6}), we obtain
\begin{multline}\label{eq:14b}
 \frac{\hbar D}{4\pi}\left(F_L-F_{\rm N}\right)\equiv \frac{\hbar D}{4\pi}\left(- A\, C_L^2 + B\, C_L^4\right) = \\
\int\limits_{0}^{R} r d r
 \bigg\{ \Delta_L^2\left(\sum\limits_{\omega_{n{\rm c}} < \Omega_{\rm D}}\frac{T_{\rm c}(H)}{\omega_{n{\rm c}}+\Omega_L}-\sum\limits_{\omega_n < \Omega_{\rm D}}\frac{T}{\omega_n+\Omega_L}\right) \\
 +\Delta_L^4 T\sum\limits_{\omega_n < \Omega_{\rm D}}\left[\frac{1}{4(\omega_n+\Omega_L)^3} +
 \frac{\Omega_L-2\hbar D\left(\frac{L-\phi_r}{r}\right)^2}{12(\omega_n+\Omega_L)^4} \right]\bigg\}
 \,.
\end{multline}
Here the second (third) line corresponds to the quadratic
(quartic) terms in $\Delta_L=C_L f_L(\phi_r)$.

Finally the amplitude $C_L$ which minimizes the above
functional $F_L = F_{\rm N} - A\, C_L^2 + B\, C_L^4$ takes the
form
$C_L^2 = A /(2 B)$, with
\begin{multline}
A = \frac{\Phi_0}{\pi \hbar D H}I_{2,0} \times\\
\Bigl[ \Psi(\omega_{L,T_{\rm c}}) - \Psi(\omega_{{\rm D},T_{\rm c}}) -
\Psi(\omega_{L,T}) + \Psi(\omega_{{\rm D},T}) \Bigr]
\end{multline}
\begin{multline}
B = \frac{1}{6(2\pi T)^3}\Bigl\{ \frac{\Phi_0I_{4,0}}{2\pi \hbar
D H} \left[ 6\pi T\zeta_3(\omega_{L,T}) + \Omega_L
\zeta_4(\omega_{L,T}) \right] \\
- \left(I_{4,1}-2L I_{4,0}+L^2
I_{4,-1}\right)\zeta_4(\omega_{L,T})\Bigr\} \ ,
\end{multline}
where $\zeta_k(a) = \sum_{n\geq 0} 1/(n+a)^k$ is the zeta
function, $\omega_{L,T} = \Omega_L/(2\pi T)+1/2$, $\omega_{{\rm D},T} =
(\Omega_{\rm D}+\Omega_L)/(2\pi T)+3/2$, and
\begin{equation}
I_{n,k} = \int_0^{\phi} f_L^n(\phi_r)\phi_r^k d\phi_r \ .
\end{equation}

\begin{figure}[t!]
\includegraphics[width=0.43\textwidth]{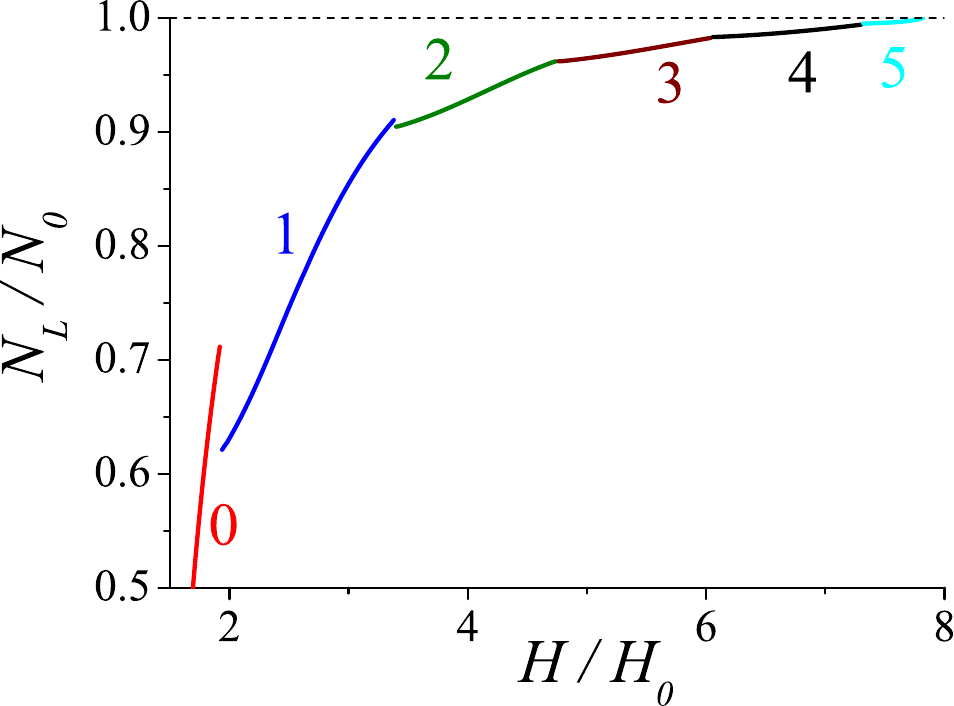}
\caption{(Color online) The normalized LDOS
$N / N_0$ at the disk edge versus the external magnetic field $H$ ($N_0$ is the
electronic density of states at the Fermi level) at $T = T_{\rm c}(H) -
0.01 T_{\rm cs}$. Here we choose $R = 4\xi_0$; $g =
0.18$.}\label{Fig3-Jump}
\end{figure}

Substituting now $\omega_n = -i \varepsilon$ in the relation
(\ref{eq:6}), one obtains the following expressions for the LDOS
valid in the first order in $\Delta_L^2(r)$ :
\begin{eqnarray} \label{eq:12}
 N_L(\varepsilon\,,r,\phi) && = \mathrm{Re} [ \mathcal{G}(\varepsilon\,,r) ]
 = \mathrm{Re} \left[ \cos\theta_L( r ) \right]\bigg|_{\omega_n = -i \varepsilon} \\
 &&\approx 1 + \frac{\Delta_L^2(r)}{2} \frac{\varepsilon^2 - \Omega_L^2}
 {\left[\, \varepsilon^2 + \Omega_L^2\, \right]^2} \nonumber
\end{eqnarray}
%
%
The transitions between different vortex states while sweeping the
magnetic field up, are visualized by abrupt changes (or jumps) in
ZBC
\cite{Nishio-PRL08,Roditchev-PRL09,Moshchalkov-PRB16}, which are
determined by the LDOS $N(\varepsilon, r, \phi)$ at
the Fermi level $\varepsilon = 0$.
 Let us consider a certain point ($T=T_{\rm s}\,,H=H_{\rm s}$)
at the phase diagram Fig.~\ref{Fig2-Phic}, where switching of the
orbital modes $L \rightleftarrows L+1$ takes place, $H_{\rm s} = H_0 \phi_L$.
Since the depairing parameters of the orbital modes $L$ and $L+1$
coincide $\Omega_L = \Omega_{L+1}$ the corresponding jump in the
LDOS $N_{L+1}-N_{L}$ at the disk edge $r = R$ can be estimated as
follows:
\begin{equation}\label{eq:13}
 N_{L+1} - N_{L}\bigg|_{\begin{subarray}{l}\epsilon=0\\ r=R\\ \phi=\phi_L\end{subarray}}
 \sim \frac{\Delta_{L}^2(R)-\Delta_{L+1}^2(R)}{2 \Omega_L^2} \ .
\end{equation}
%
Figure~\ref{Fig3-Jump} shows the magnetic field
dependence of the normalized LDOS at the disk edge. The
transitions between different vortex states ($L \to L+1$) are
accompanied by the abrupt reduction in LDOS at the
disk edge while sweeping the magnetic field up. Similar jumps of
the LDOS, which are attributed to the entrance of a vortex inside
the disk, have been observed in 
measurements of the
normalized ZBC on $\mathrm{Pb}$ nano-island~\cite{Roditchev-PRL09}
and $\mathrm{MoGe}$ nanostructures~\cite{Moshchalkov-PRB16}.
%
%

\subsection{An arbitrary magnetic field: $0 < H < H_{\rm c2}$}
As a next step, we analyze the conductance behavior as a function of magnetic field and temperature
at arbitrary magnetic fields, $0 < H < H_{\rm c2}$.
The Usadel equations (\ref{eq:1a}~-~\ref{eq:4}) have been solved numerically for different vorticities which allowed us to calculate and compare
the values of the free energy.
\begin{figure}[t]
\includegraphics[width=0.45\textwidth]{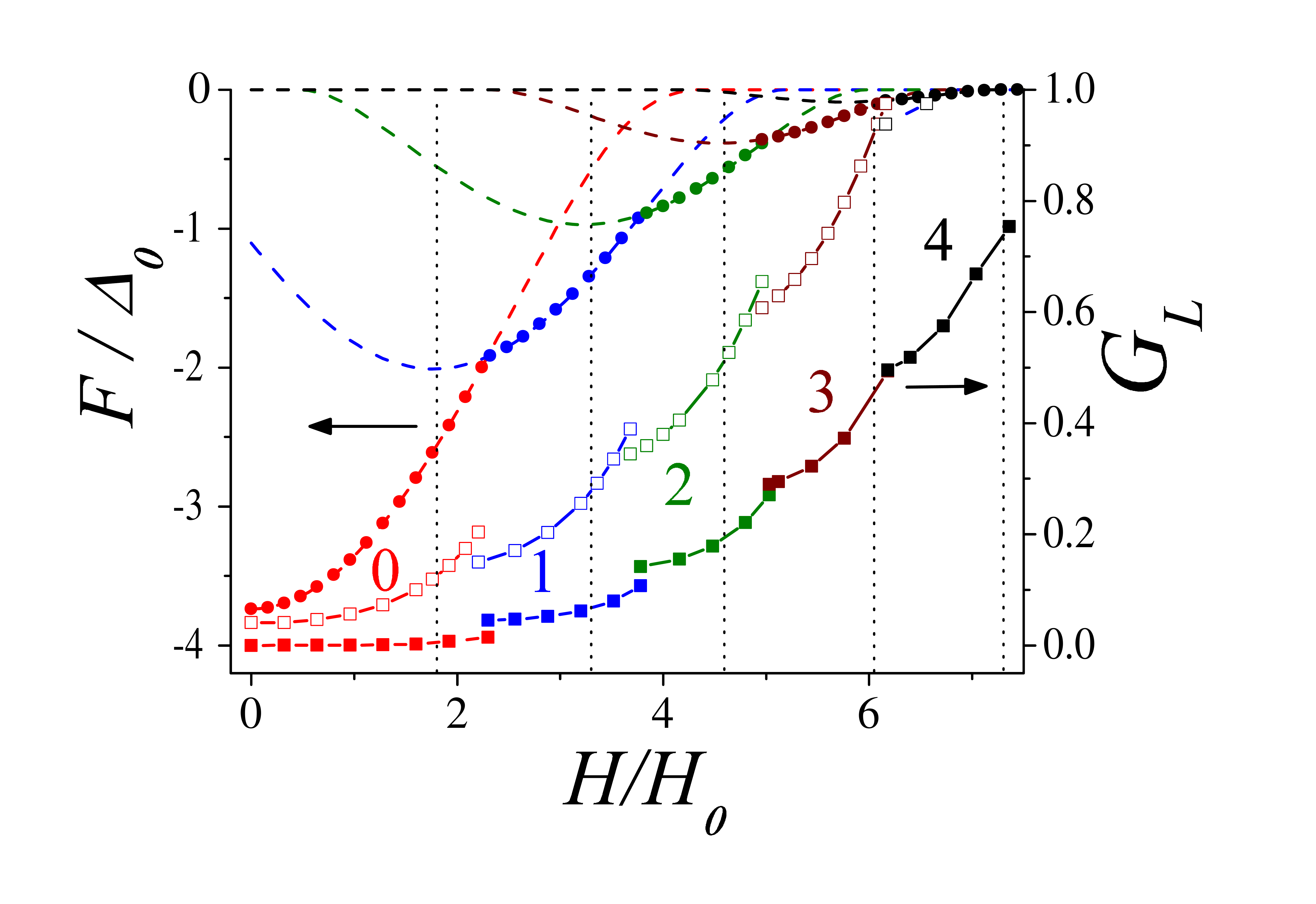}
\caption{(Color online) The dependence of the free energy
$F(\phi)$ (\ref{usadelfull}) (symbol $\bullet$) and the normalized
zero bias conductance (ZBC) $G_L(0\,,R\,,\phi)$ (\ref{eq:16})
at the disk edge for the temperatures $T = 0. 1 T_{\rm cs}$ (symbol~$\blacksquare$) and
$T = 0. 2 T_{\rm cs}$ (symbol~$\square$)
on the magnetic flux $\phi
= H / H_0$ across the SC disk of the radius $R = 4 \xi_0$. The
dashed lines show the dependence $F_L(\phi)$ for fixed vorticity
$\mathrm{L=0 \div 4}$. The numbers near the curves denote the
corresponding values of vorticity $L$. Vertical dotted lines $H /
H_0 = \phi_L$ correspond to the switching of the orbital modes in
the critical temperature $T_{\rm c}$, shown in Fig.~\ref{Fig1}.}
\label{Fig4-R04T01}
\end{figure}
%
Figure~\ref{Fig4-R04T01} shows the magnetic field dependence of
the free energy $F$ (\ref{usadelfull}) and the zero bias conductance
$G_L(0\,,R\,,\phi)$ (\ref{eq:16}) at the Fermi level for a
small disk radius $R = 4 \xi_0$ and two temperatures $T = 0. 1 T_{\rm cs}$
and $T = 0. 2 T_{\rm cs}$.
All three curves illustrate the
switching between the states with different vorticities $L = 0
\div 4$, which are similar Little-Parks-like switching of the
critical temperature $T_{\rm c}(H)$, Fig.~\ref{Fig1}. Sequential entries of vortices
produce a set of branches $F_L$ with different vorticity $L$ on the
$F(H)$ and $dI/dV(H)$ curves.
The transitions between different vortex states are accompanied by
an abrupt change in the ZBC, which is attributed to the entry/exit
of a vortex inside the disk while sweeping the magnetic field.
We observe the Meissner state when
the total vorticity $L=0$ for $H \lesssim H_{{\rm s}0}=2.24 H_0$, and a
single-vortex state $L=1$ in the field range $H_{{\rm s}0} \lesssim H
\lesssim H_{{\rm s}1} = 3.84 H_0$.
In the Meissner state the ZBC is suppressed and
spatially homogeneous: the ZBC value at the disk edge is slightly
higher then ZBC value in the center. In the increasing magnetic field the
gap in the tunneling spectra gradually fills with the quasiparticle
states. This effect is more pronounced near the disk edge where
the screening superconducting currents have higher density. The
smooth evolution of ZBC continues till $H / H_0 \simeq 2.24$ where
it is interrupted by a vortex entry.
At higher fields $H > H_{{\rm s}1}$ the
multivortex states $L = 2 \div 4$ become energetically favorable.
\rev{
}
Note, that the field values $H_{{\rm s}L}$ at which the
jumps in vorticity ($L \to L+1$) occur are always larger than the values
$H_0\,\phi_L$ found from the calculations of the critical
temperature behavior.
%
\begin{figure}[t]
\includegraphics[width=0.45\textwidth]{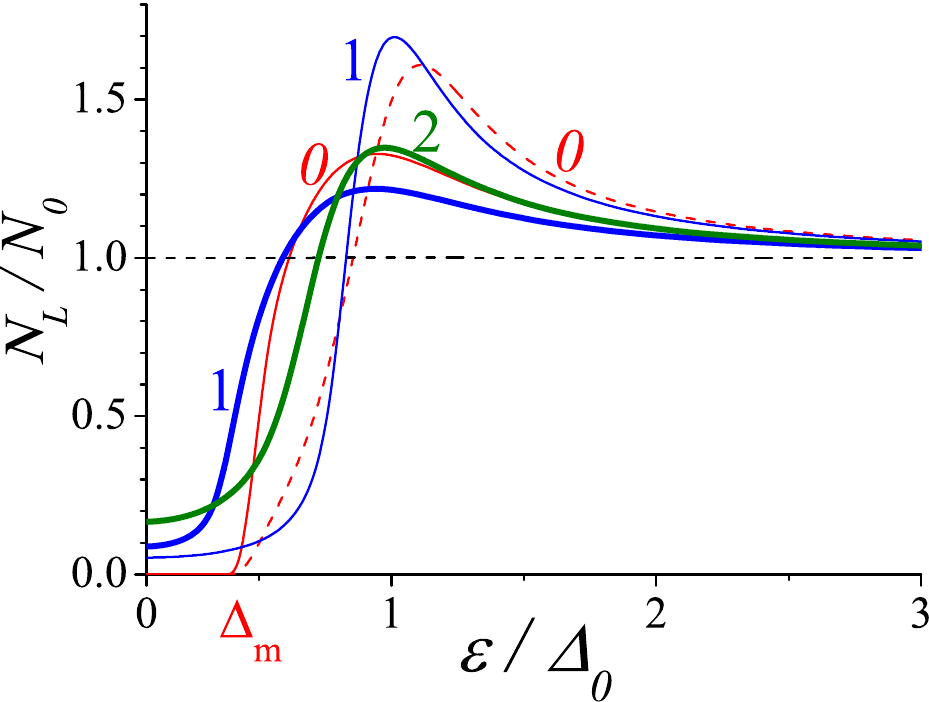}
\caption{(Color online) Evolution of the spatially resolved LDOS
$N(\varepsilon, r, \phi)$ in the disk center $r=0$ (dashed
lines) and the disk edge $r=R$ (solid lines) in the magnetic
field: thin lines -- $H / H_0 \simeq \mathrm{2.24}$; bold lines $H
/ H_0 \simeq \mathrm{3.84}$ ($R = 4\,\xi_0$, $T = \mathrm{0. 1}
T_{\rm cs} $). The numbers near the curves denote the corresponding
values of vorticity $L$. } \label{Fig6-R04T01}
\end{figure}
%

Figure \ref{Fig4-R04T01} illustrates an important point noted in Introduction, i.e.,
the temperature crossover between different regimes in the behavior of the conductance vs magnetic field.
Indeed, one can clearly see that the change in temperature from $0.1 T_{\rm cs}$ to $0.2 T_{\rm cs}$ is accompanied by the
change of the direction of jumps in the dependence of zero bias conductance vs magnetic field.
The upward jumps
in conductance for the lower temperature, $T<T^*(R)$, can be associated with the core dominated regime $e^{-E_{\rm g}/T}<e^{-R/d_L}$, see Fig.~\ref{Fig0}, when the conductance increases with the increase
in the number of vortices trapped in the center of the sample and therefore in the parameter $d_L$.
The downward jumps in conductance at higher temperatures, $T>T^*(R)$, are caused by the increase in the (soft) spectral gap value $E_{\rm g}$ at the sample edge as the vortex enters, which should
result in the suppression of the subgap conductance $G_L \sim e^{-E_{\rm g}/T}$. 
The change in vorticity $L \to L+1$ in this case results in the decrease of the
screening current density and the corresponding enhancement of superconductivity at the edge of the disk.
Assuming the crossover temperature $T^*(R)$ to be in the interval $0.1 T_{\rm cs}<T^*(R)<0.2 T_{\rm cs}$ and taking $R=4\xi_0$ one can estimate the value
$E_{\rm g}\sim 0.8\Delta_0$ which is in good agreement with the behavior of the energy dependence of the local density of states in Fig.~\ref{Fig6-R04T01}.
Indeed, the position of the maximal slope of the energy dependence of the density of states roughly gives the value of the minigap at the edge:
$E_{\rm g} \lesssim 0.8\Delta_0$.

Figure~\ref{Fig6-R04T01} also illustrates the switching between the states with hard and soft gaps
with the increase in the magnetic field.
In the Meissner state ($H <H_{{\rm s}0}$) the hard minigap $\Delta_{\rm m}$ in the spectrum exists
($N(\varepsilon < \Delta_{\rm m}, r, \phi) = 0$) till the first vortex entry.
%
The
density of states in the center of the disk $N(\varepsilon,
\mathrm{0}, \phi)$ is equal to the electronic density of states
at the Fermi level $N_0$ for any vortex state $L \ge 1$,
indicating a full suppression of the spectral gap in the disk
center due to the vortex entry. At the same time, at the edge of
the disk the superconductivity survives though the gap
becomes soft, $0 < N(0, R, \phi) < N_0$.
%

\begin{figure}[t]
\includegraphics[width=0.45\textwidth]{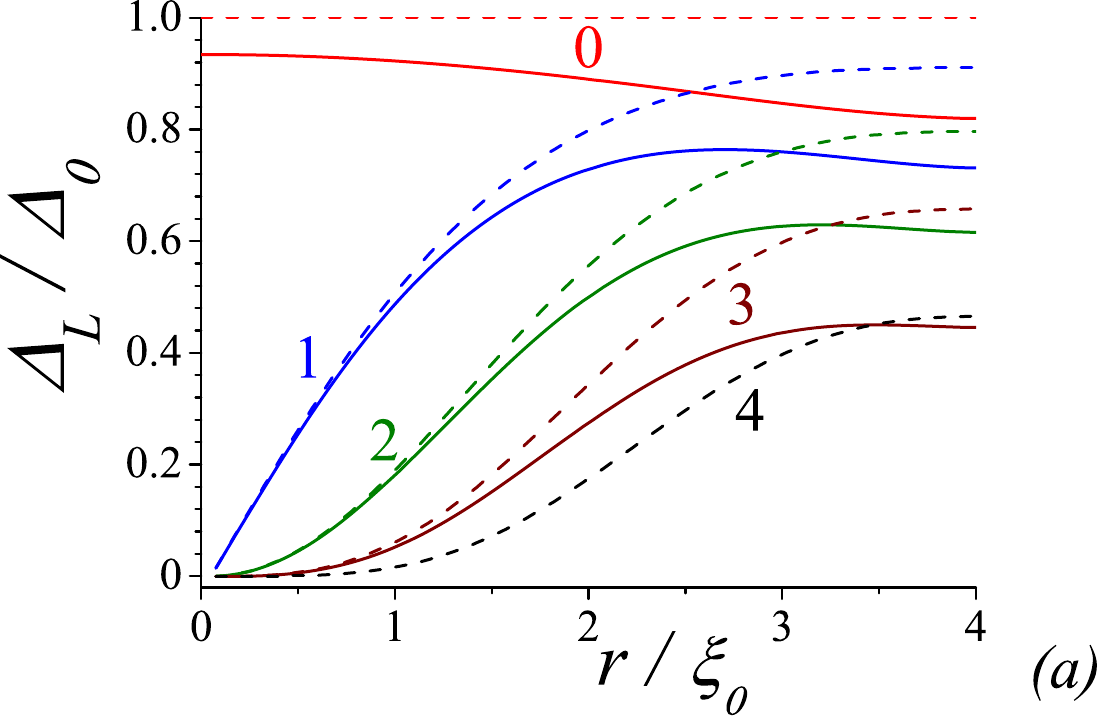}
\includegraphics[width=0.45\textwidth]{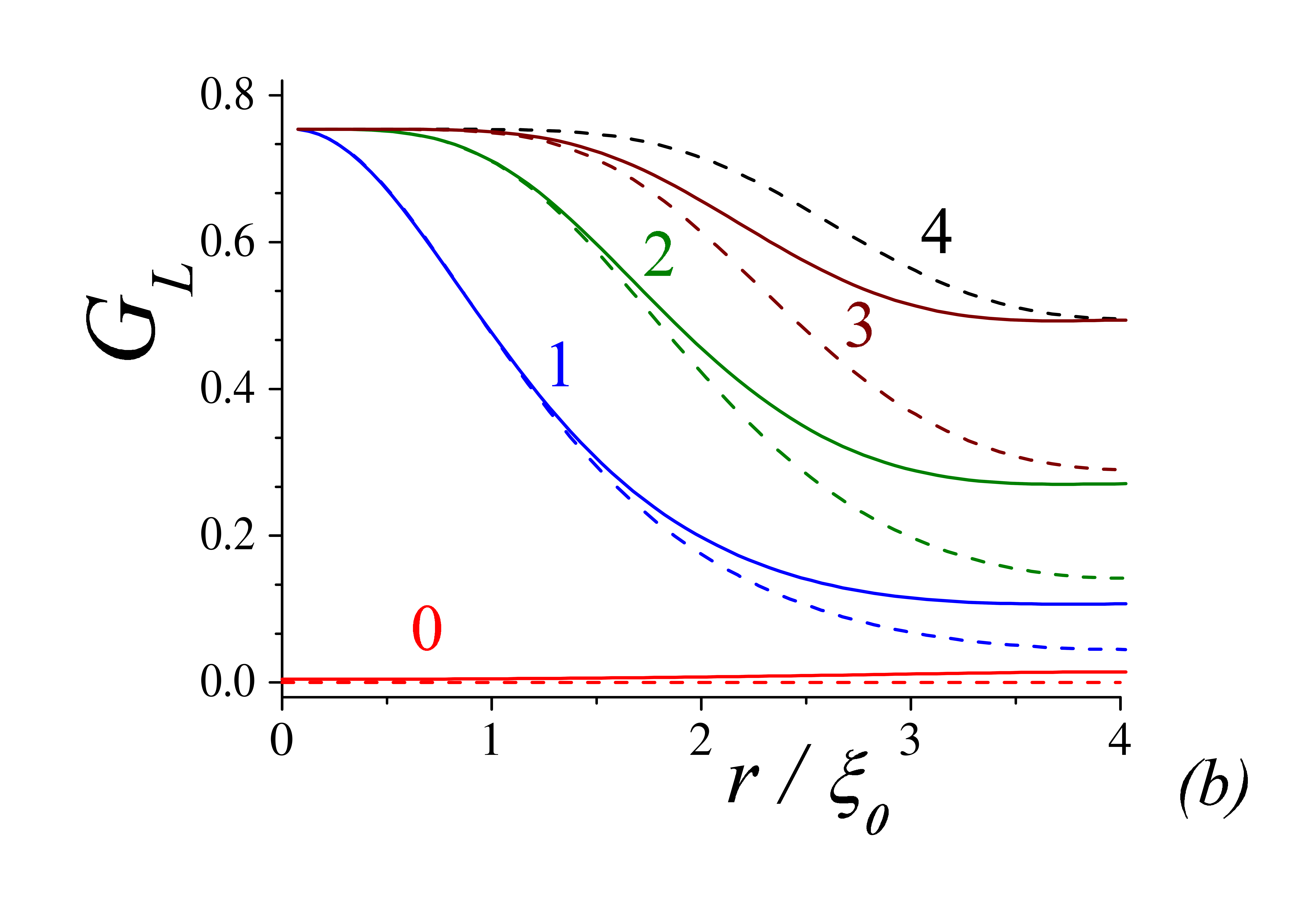}
\caption{(Color online) The radial dependence of the SC order parameter
$\Delta_L$ ($a$) and ZBC $G_L(0, R, \phi)$ ($b$) for minimal
(dashed line) and maximal (solid line) allowed values of the
magnetic field $H$ for the orbital mode $L$: $L=0$ -- $H / H_0 =
\mathrm{0\,,2.24}$; $L=1$ -- $H / H_0 = \mathrm{2.24\,,3.84}$;
$L=2$ -- $H / H_0 = \mathrm{3.84\,,4.96}$; $L=3$ -- $H / H_0 =
\mathrm{4.96\,,6.08}$; $L=4$ -- $H / H_0 = \mathrm{6.08}$ ($R =
4\,\xi_0$, $T = \mathrm{0. 1} T_{\rm cs} $). The numbers near the
curves denote the corresponding values of vorticity $L$. }
\label{Fig5-R04T01}
\end{figure}

Figure~\ref{Fig5-R04T01} presents the radial distributions of the
SC order parameter $\Delta_L(r)$ and the ZBC $dI / dV$ at $T = 0. 1\,T_{\rm c}$ for
different values of the magnetic field $H$ corresponding to the
switching between the states with different vorticity $L$. The
profiles of ZBC in multiquantum vortices $L > 1$ reveal a plateau
near the vortex center, which can be considered as a hallmark of the
multiquantum vortex formation in dirty mesoscopic superconductors
\cite{Roditchev-PRL11,Roditchev-PRL09,Silaev-JPCM13}.
%
%

%
%
%
%

The electronic properties of the vortex states look to be rather different if the radius of the
disk $R$ is much larger than the coherence length $\xi_0$. In
this case the core of a multiquantum vortex does not extend to the
edge of the disk, and quasiparticles in the vortex core remain well
localized near the disk center. Clearly, in this case the temperature crossover between the core-dominated and edge-dominated regimes
accompanied by the change in the direction of the jumps in the local ZBC at the sample edge becomes much more difficult to observe
due to the exponentially small values of the factors $\exp(-R/d_L)$ and $\exp(-E_{\rm g}/T)$ near the crossover.
%
%

\section{Conclusions}\label{SumUp}
To sum up, we have analyzed the behavior of the LDOS
$N(\varepsilon\,,r\,,H)$ and conductance on an external magnetic field $H$
in a mesoscopic superconducting disk on the basis of Usadel
equations. We have demonstrated that transitions between the
superconducting states with different vorticities provoke abrupt
changes (jumps) in the local zero bias conductance $d I / d V$ at
the edge of the disk. These jumps of the ZBC are attributed to the
entry/exit of vortices while sweeping the magnetic
field. The transitions between different vortex states can be
accompanied both by the decrease and increase in the ZBC while sweeping
the magnetic field up. The direction of jumps in ZBC attributed to
the vortex entry depends on the disk radius
$R$ and the temperature $T$ and is determined by two opposite in sign
contributions to conductance: (i)~the entrance of a vortex into the disk is
accompanied by the reduction of the supercurrents flowing along the
sample edge and, thus, improves superconductivity at the edge; (ii)~the entrance of a vortex
increases the number of subgap quasiparticle states in the multiquantum vortex core
which provide an additional contribution to the conductance because of the quasiparticle tunneling
between the vortex core and the sample edge.
To the best of our knowledge, the systematic experimental analysis of the direction of the ZBC jumps
has not been done yet.
However, these measurements can provide an additional information about the soft gap value governing
not only the contribution to the tunneling transport, but also the one to the thermal relaxation mechanisms (see, e.g., \cite{Khaymovich-NatCom16,Nakamura17})
and also about the classical-to-quantum interplay in quasiparticle tunneling in mesoscopic superconducting samples.
These results are directly related to the quantitative characterization of the
quasiparticle traps appearing in the Meissner and
vortex states of superconductors (see, e.g., \cite{Peltonen2011,Nsanzineza2014,Wang2014,Vool2014,Woerkom2015,Khaymovich-NatCom16,Nakamura17}),
especially in the different types of single-electron sources based on hybrid superconducting junctions and working far from equilibrium~\cite{Khaymovich-NatCom16,Nakamura17,parity_NISIN2015,Khaymovich_Basko2016,vanZanten2016}.

\acknowledgments This work was supported, in part, by the Russian
Foundation for Basic Research under Grant No. 17-52-12044.
In the part concerning the LDOS numerical calculations, the
work was supported by Russian Science Foundation (Grant No.
17-12-01383). IMK acknowledges the support of the German Research
Foundation (DFG) Grant No. KH~425/1-1.

\appendix

\section{Solution of Eq.~(\ref{eq:1b}) via the Kummer's functions}
\label{App1}
Substituting the expression of the order parameter $\Delta_L(r) =
\theta_L(r)\left(\omega_n+\Omega_L\right)$, \eqref{eq:6}, into
Eq.~\eqref{eq:1b} and rewriting the latter in terms of the
renormalized flux $\phi_r = \pi r^2 H / \Phi_0$,
\begin{gather}
 \frac{d}{d \phi_r} \left( \phi_r\,\frac{d \theta_L}{d \phi_r} \right) - \frac{( L - \phi_r)^2}{4 \phi_r}\theta_L
 +\frac{\Phi_0 \Omega_L}{2\pi \hbar D H} \theta_L = 0
\end{gather}
one can easily obtain the equation for the function $W(\phi_r)$ defined as
\begin{gather}\label{App:theta_L}
\theta_L(r)= e^{-\phi_r/2}\phi_r^{|L|/2}W(\phi_r) \
\end{gather}
\begin{equation}\label{eq:A4}
 \phi_r\, \frac{d^2 W}{d \phi_r^2} + ( b_L - \phi_r )\, \frac{d W}{d \phi_r} - a_L\, W
 = 0 \ ,
\end{equation}
with the parameters $a_L$ and $b_L$ given by
\begin{gather}
a_L = \frac{1}{2}\, \left( | L | - L + 1 - \frac{\Phi_0 \Omega_L}{\pi \hbar D H} \right)\,, \quad b_L = | L | + 1 \ .
\end{gather}

The solution of Eq.~(\ref{eq:A4}) in the region $r \le R$ is
confluent hypergeometric function of the first kind (Kummer's
function), $W = K(a_L,\,b_L,\,\phi_r)$, which after substitution to
the expression \eqref{App:theta_L} for $\theta_L(r)$ gives the
result \eqref{eq:4ab} from the main text. 

%
%
%

%

\end{document}